\documentclass[a4paper,useAMS,usenatbib]{mn2e}

\usepackage{amsmath}
\usepackage{amssymb}
\usepackage{graphicx} 
\usepackage{sansmath}   
\usepackage{flafter}
\usepackage{float}
\usepackage{fixltx2e}
\usepackage{url}

\newcommand{\ddp}[2]{\frac{\upartial #1}{\upartial #2}}

\newcommand{\lddp}[2]{\upartial #1/\upartial #2}

\newcommand{\eexp}{\mathrm{e}}
\newcommand{\bJ}{\boldsymbol{J}}
\newcommand{\bth}{\boldsymbol{\theta}}

\newcommand{\Rb}{R_{\mathrm{b}}}

\newcommand{\Oms}{\Omega_\mathrm{s}}

\newcommand{\Omb}{\Omega_\mathrm{b}}

\newcommand{\ROLR}{R_{\mathrm{OLR}}}

\newcommand{\Kpc}{\,\mathrm{kpc}}

\newcommand{\kmsec}{\,\mathrm{km}\,\mathrm{s}^{-1}}

\newcommand{\vc}{v_{\mathrm{c}}}

\newcommand{\de}{\mathrm{d}}

\newcommand{\Rg}{R_\mathrm{g}}

\newcommand{\RNum}[1]{\uppercase\expandafter{\romannumeral #1\relax}}

\newcommand{\Ec}{E_\mathrm{c}}

\newcommand{\ldd}[2]{\de #1/\de #2}

\newcommand{\pare}[1]{\left(#1\right)}
\newcommand{\paresq}[1]{\left[#1\right]}
\newcommand{\parec}[1]{\left\{#1\right\}}

\newcommand{\img}{\mathrm{i}}
\newcommand{\Rep}{\operatorname{Re}}

\newcommand{\Eq}[1]{Eq.~(\ref{#1})}
\newcommand{\Eqs}[2]{Eqs.~(\ref{#1})-(\ref{#2})}

\newcommand{\Fig}[1]{Fig.~\ref{#1}}

\newcommand{\Phia}{\Phi_\mathrm{a}}

\newcommand{\App}[1]{Appendix~\ref{#1}}

\newcommand{\sgn}{\mathrm{sgn}}

\newcommand{\Js}{J_{\mathrm{s}}}
\newcommand{\Ja}{J_{\mathrm{a}}}
\newcommand{\Jf}{J_{\mathrm{f}}}
\newcommand{\Jp}{J_{\mathrm{p}}}
\newcommand{\Jh}{J_{\mathrm{h}}}
\newcommand{\Jsres}{J_{\mathrm{s,res}}}

\newcommand{\ths}{\theta_{\mathrm{s}}}
\newcommand{\thf}{\theta_{\mathrm{f}}}
\newcommand{\thp}{\theta_{\mathrm{p}}}

\newcommand{\oH}{\overline{H}}
\newcommand{\Vp}{V_\mathrm{p}}

\newcommand{\Ep}{E_\mathrm{p}}

\newcommand{\alphab}{\alpha_\mathrm{b}}

\newcommand{\sn}{\mathrm{sn}}
\newcommand{\cn}{\mathrm{cn}}
\newcommand{\dn}{\mathrm{dn}}
\newcommand{\ftr}{f_\mathrm{tr}}
\newcommand{\fcirc}{f_\mathrm{circ}}

\def\apjl{ApJ}%
\def\apjs{ApJS}%
\def\ao{Appl.~Opt.}%
\def\aap{A\&A}%
\title[Distribution functions for trapped orbits]{Distribution
  functions for resonantly trapped orbits in the Galactic disc}

\author[G.~Monari~et~al.]{Giacomo Monari$^1$\thanks{E-mail:
    giacomo.monari@fysik.su.se}, Benoit Famaey$^2$, Jean-Baptiste Fouvry$^{3}$\thanks{Hubble fellow} and James Binney$^4$\\
$^1$ The Oskar Klein Centre for Cosmoparticle Physics, Department of Physics, Stockholm University, AlbaNova, 10691 Stockholm, Sweden\\
$^2$ Universit\'e de Strasbourg, CNRS UMR 7550, Observatoire astronomique de Strasbourg, 11 rue de l'Universit\'e, 67000 Strasbourg, France \\
$^3$ Institute for Advanced Study, Einstein Drive, Princeton, NJ 08540, USA \\
$^4$ Rudolf Peierls Centre for Theoretical Physics, Keble Road, Oxford OX1 3NP, UK}

\date{Accepted XXX. Received YYY; in original form ZZZ}

\pubyear{2017}

\begin{document}
\label{firstpage}
\pagerange{\pageref{firstpage}--\pageref{lastpage}}
\maketitle

\begin{abstract}
The present-day response of a Galactic disc stellar population to a
non-axisymmetric perturbation of the potential has previously been
computed through perturbation theory within the phase-space
coordinates of the unperturbed axisymmetric system. Such an {\it
  Eulerian} linearized treatment however leads to singularities at
resonances, which prevent quantitative comparisons with data. Here, we
manage to capture the behaviour of the distribution function (DF) at a
resonance in a {\it Lagrangian} approach, by averaging the Hamiltonian
over fast angle variables and re-expressing the DF in terms of a new
set of canonical actions and angles variables valid in the resonant
region. We then follow the prescription of Binney~(2016), assigning to
the resonant DF the time average along the orbits of the axisymmetric
DF expressed in the new set of actions and angles. This boils down to
phase-mixing the DF in terms of the new angles, such that the DF for
trapped orbits only depends on the new set of actions. This opens the
way to quantitatively fitting the effects of the bar and spirals to
Gaia data in terms of distribution functions in action space.
\end{abstract}


\section{Introduction}

The optimal exploitation of the next data releases of the Gaia mission
\citep{Gaia} will necessarily involve the construction of a fully
dynamical model of the Milky Way. Rather than trying to construct a
quixotic full {\it ab initio} hydrodynamical model of the Galaxy,
which would never be able to reproduce all the details of the Gaia
data, a promising approach is to rather construct a multi-component
phase-space distribution function (DF) representing each stellar
population as well as dark matter, and to compute the potential that
these populations jointly generate
\citep[e.g.,][]{BinneyPiffl2015}. To do so, one can make use of Jeans
theorem, constraining the DF of an equilibrium configuration to depend
only on three integrals of motion. Choosing three integrals of motion
which have canonically conjugate variables, allows us to express the
Hamiltonian $H_0$ in its simplest form, i.e. depending only on these
three integrals.  Such integrals are called the ``action variables"
$\bJ$, and are new generalized momenta having the dimension of
velocity times distance, while their dimensionless canonically
conjugate variables are called the ``angle variables" $\bth$, because
they are usually normalized such that the phase-space position is
$2\pi$-periodic in them \citep[e.g.][]{BT2008}. In absence of
perturbations, these angles evolve linearly with time, $\bth(t)=\bth_0
+ \boldsymbol{\Omega} t$, where
$\boldsymbol{\Omega}(\bJ)\equiv\lddp{H_0}{\bJ}$ is the vector of
fundamental orbital frequencies. In an equilibrium configuration, the
angle coordinates of stars are phase-mixed on orbital tori that are
defined by the actions $\bJ$ alone, and the phase-space density of
stars $f_0(\boldsymbol{J}) \de^3\boldsymbol{J}$ corresponds to the
number of stars $\de N$ in a given infinitesimal action range divided
by $(2\pi)^3$. In an axisymmetric configuration, the action variables
can simply be chosen to be the radial, azimuthal and vertical actions
respectively. By constructing DFs depending on these action variables,
the current best axisymmetric models of the Milky Way have been
constructed \citep{BinneyCole}.

The Milky Way is however not axisymmetric: it harbours both a bar
\citep{deVaucouleurs1964,Binney1991,Binney1997,Wegg2015,Monari2017a,Monari2017b}
and spiral arms, the exact number, dynamics and nature of which are
still under debate \citep{SellwoodCarlberg2014,Grand2015}. Whilst
\citet{Trick2017} showed that spiral arms might not affect the
axisymmetric fit, the combined effects of spiral arms and the central
bar of the Milky Way are clearly important observationally
\citep[e.g.,][]{McMillan2013,Bovy2015}. Hence, non-axisymmetric
distribution functions are needed to pin down the present structure of
the non-axisymmetric components of the potential, which have enormous
importance as drivers of the secular evolution of the disc
\citep{Fouvry,Fouvry2015,Aumer2016,Aumer2017}.

A recent step \citep{Monari2016} has been to derive from perturbation
theory explicit distribution functions for present-day snapshots of
the disc as a function of the actions and angles of the unperturbed
axisymmetric system.  This work, which is an {\it Eulerian} approach
to the problem posed by non-axisymmetry, has allowed us to probe the
effect of stationary spiral arms in three spatial dimensions, away
from the main resonances. In particular, the moments of the perturbed
distribution function describe ``breathing" modes of the Galactic disc
in perfect accordance with simulations \citep{Monari2016}. Such a
breathing mode might actually have been detected in the extended Solar
neighbourhood \citep{Williams2013}, but with a larger amplitude,
perhaps because the spiral arms are transient.  Although such an {\it
  Eulerian} treatment has also been used to gain {\it qualitative}
insights on the effects of non-axisymmetries near resonances
\citep{Monari2017a}, no quantitative assessments can be made with such
an approach, because the linear treatment diverges at resonances
\citep[the problem of small divisors, ][]{BT2008}.

In the present contribution, we solve this problem by developing the
{\it Lagrangian} approach to the impact of non-axisymmetries at
resonances.  The basic idea is to model the deformation of the orbital
tori outside of the trapping region, and to construct new tori,
complete with a new system of angle-action variables, within the
trapping region \citep{Kaasalainen1994}.  Finally, following
\citet{Binney2016} we populate the new tori by phase-averaging the
unperturbed distribution function over the new tori.

In Section~2, we present some examples of trapped and untrapped
orbits. In Section~3, we summarise the standard approach to a
resonance, namely to make a canonical transformation to fast and slow
angles and actions, and to replace the real Hamiltonian by its average
over the fast angles \citep{Arnold}. Under this averaged Hamiltonian
the slow variables have the dynamics of a pendulum.  In Section~4, we
introduce the pendulum's angles and actions.  In Section~5, we discuss
how to build the distribution function using the newly introduced
pendulum angles and actions, both inside and outside the zone of
trapping at resonances. In Section~6, we present the form of the
distribution functions in velocity space, in cases of astrophysical
interest related to the Galactic bar. We conclude in Section~7.

\section{The bar and trapped orbits}

Let us consider orbits in the Galactic plane, and let $(R,\phi)$ be
the Galactocentric radius and azimuth. The logarithmic potential,
corresponding to a flat circular velocity curve $\vc(R)=v_0$, is a
rough but simple representation of the potential of the Galaxy. In a
formula,
\begin{equation}\label{eq:phi0}
  \Phi(R)=v_0^2\ln(R/R_0),
\end{equation}
with $R_0$ the distance of the Sun from the Galactic center. Motion in this planar
axisymmetric potential admits actions $J_\phi$, which is simply the angular
momentum about the Galactic centre, and  $J_R$, which quantifies the extent
of radial oscillations.

Let the axisymmetric potential be perturbed by a non axisymmetric component
\begin{equation}\label{eq:Fmode}
  \Phi_1(R,\phi,t)=\Rep\parec{\Phia(R)\eexp^{\img m(\phi -\Omb t)}},
\end{equation}
where $\Omb$ is the pattern speed. We will specialise to the bar adopted by
\cite{Dehnen2000} and \cite{Monari2017a}, so we set $m=2$ and adopt
\begin{equation}
  \Phia(R)=-\alphab\frac{R_0^5 \Omega_0^2}{3 \Rb^3}
  \times \left\{
  \begin{array}{l l}
    (R/\Rb)^{-3}  & \quad \text{} R \geq \Rb,\\
    2 -(R/\Rb)^{3} & \quad \text{} R < \Rb,
  \end{array} \right.
\end{equation}
where $\Omega_0\equiv\Omega(R_0)$ is the circular frequency at the
solar radius, $\Rb$ is the length of the bar and $\alphab$ represents
the maximum ratio between the radial force contributed by the bar and
the axisymmetric background at the Sun \citep[see
  also][]{Monari2015,Monari2016b}. We further set $R_0=8.2\Kpc$,
$v_0=243\kmsec$, $\Rb=3.5\Kpc$, and $\alpha=0.01$,

\begin{figure*}
  \centering 
  \includegraphics[width=0.95\columnwidth]{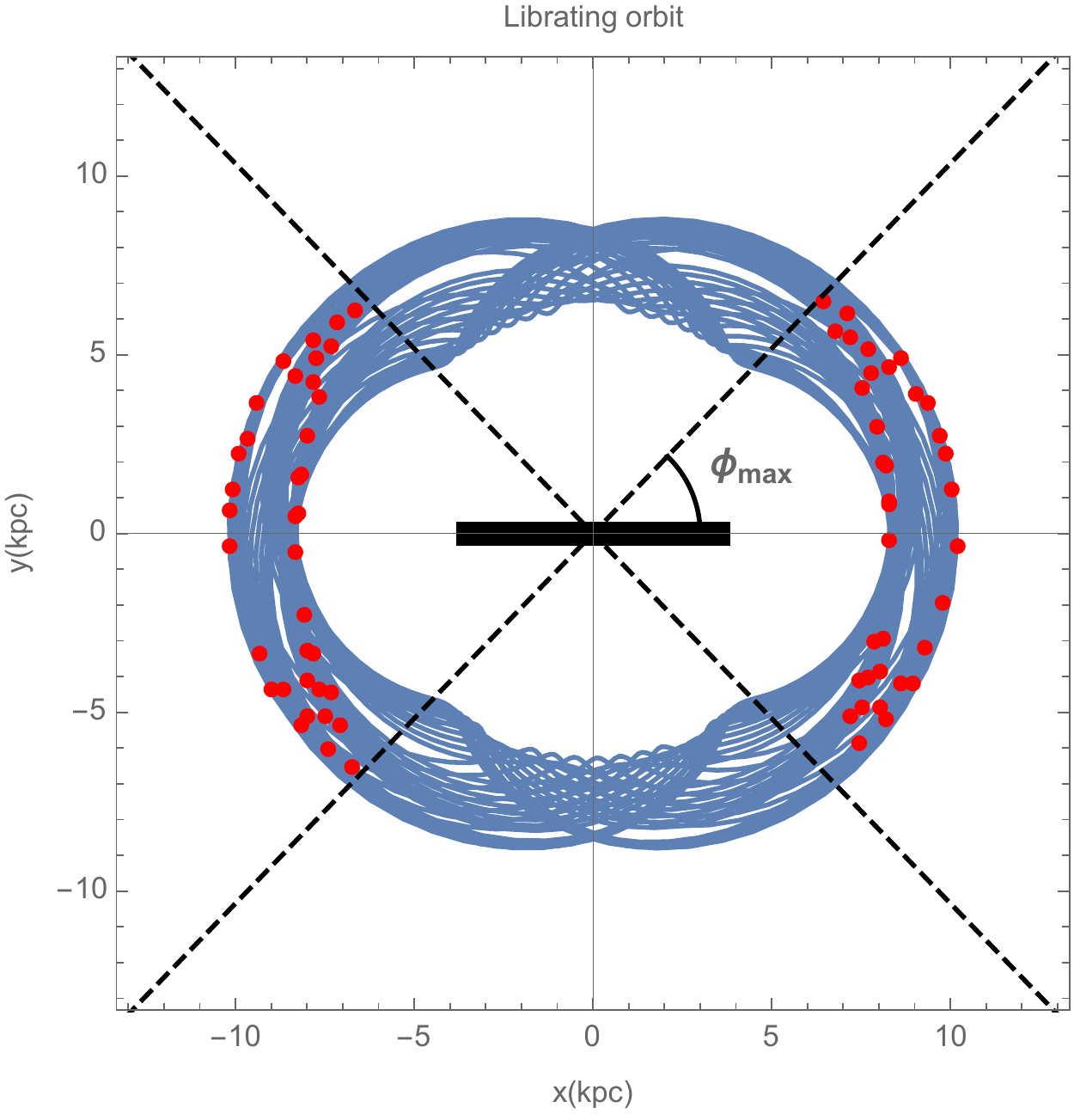}
  \includegraphics[width=0.95\columnwidth]{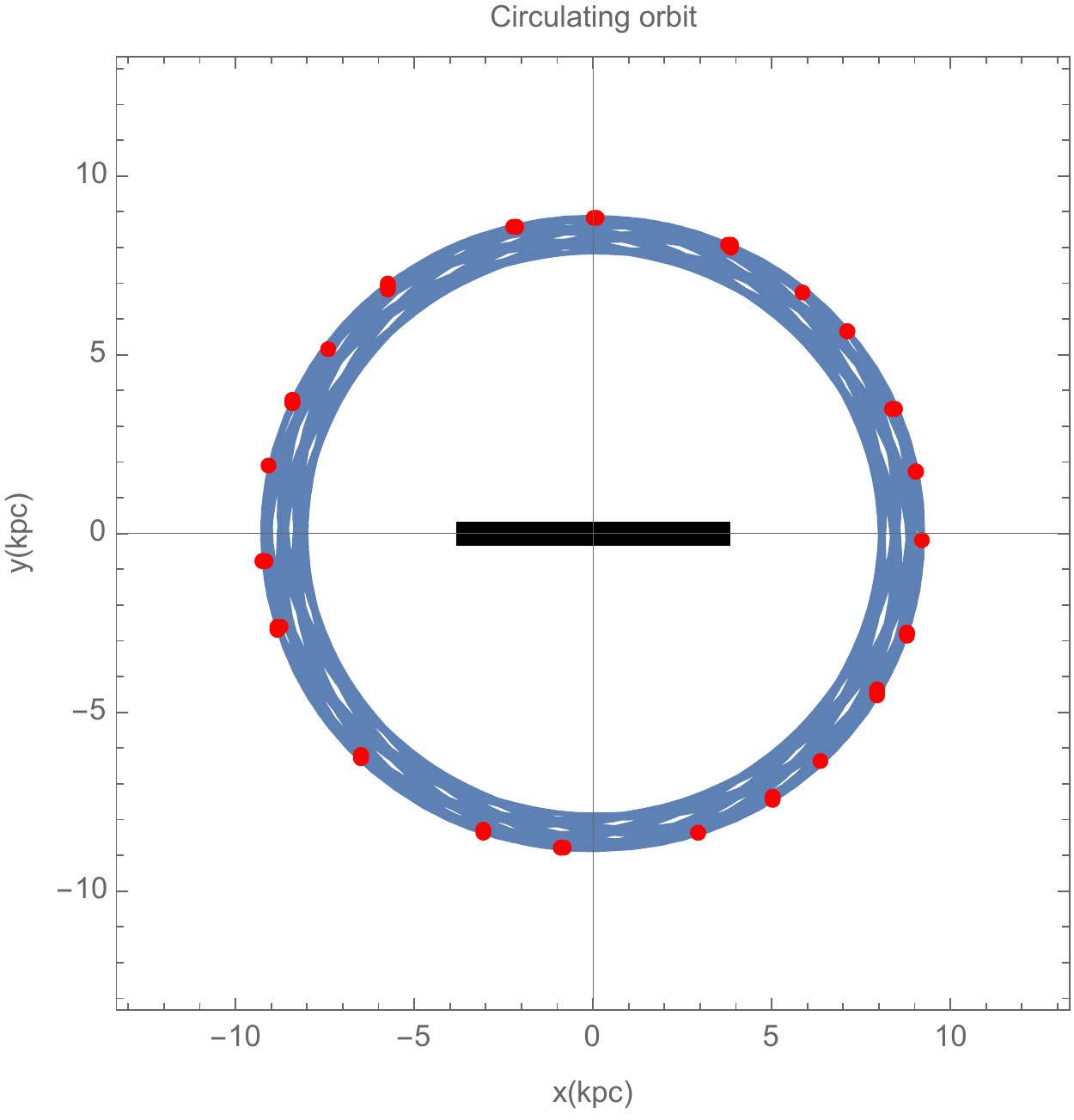}
  \caption{Two orbits in the Galactic potential described in this
    work, shown in the reference frame corotating with the bar. Here,
    the pattern speed is chosen to be $\Omb=1.8\Omega_0$. The red
    points correspond to the position of the relative apocentres. The
    thick horizontal line represents the long axis of the bar. The bar
    rotates counterclockwise. Left panel: an orbit trapped to the
    outer Lindblad resonance, where the angle $\phi_\mathrm{max}$
    (dashed lines) represents the maximum extension of the apocentres
    excursions. Right panel: a circulating orbit.}
\label{fig:trap_orb}
\end{figure*}

\Fig{fig:trap_orb} shows for $\Omb =1.8\Omega_0$ two orbits in the
reference frame corotating with the bar. The red points in this plot
correspond to the relative apocentres of the orbit.\footnote{While in
  an axisymmetric potential the apocentres of the orbits in the
  Galactic plane are always at the same distance $R$ from the Galactic
  centre, this is not the case in a non-axisymmetric potential like
  the one used in this work. Therefore, the plotted points correspond
  to relative apocentres.} In the left panel of \Fig{fig:trap_orb},
the azimuths of the apocentre points do not cover the whole
$[0,2\upi]$ range, but rather oscillate within the interval
$[-\phi_\mathrm{max},\phi_\mathrm{max}]$. The orbit is said to be
``trapped'' at a resonance, in this case the outer Lindblad resonance
(see below), and the angle of the apocentres, the ``precession
angle'', is said to ``librate''. For comparison, the right panel of
\Fig{fig:trap_orb} shows an orbit for which the precession angle
covers the whole $[0,2\upi]$ range. This orbit is said to
``circulate''. In the following, we give a quantitative description of
these two types of orbits, using perturbation theory.

\section{Reduction to a pendulum}

We will hereafter work within the epicyclic approximation
\citep{BT2008}, in which radial oscillations are harmonic with angular
frequency $\kappa(R)$, so the radial action $J_R = E_R / \kappa$,
where $E_R=E-\Ec$ is the energy of these oscillations ($\Ec$ being the
energy of a circular orbit with the same angular momentum
$J_\phi$). Using the formulae reported in \cite{Dehnen1999} and
\cite{Monari2017a}, we can then relate the coordinates of the trapped
orbit to the angles $\theta_R$ and $\theta_\phi$ of the unperturbed
system.  We can also rewrite the perturbing potential in the Galactic
plane in terms of actions and angles as a Fourier series
\begin{equation}\label{eq:pot_bar}
  \Phi_1 (J_R, J_\phi, \theta_R, \theta_\phi) =
  \Rep\Bigg\{\sum_{j=-1}^{1} c_{j m} \eexp^{\img
    \paresq{j\theta_R+m\pare{\theta_\phi- \Omb t}}}\Bigg\},
\end{equation}
with
\begin{align}\label{eq:cR}
  c_{j m}(J_R,J_\phi)&\equiv \Bigg[\delta_{j 0}+\delta_{|j|1}\frac{m}{2}\sgn(j)
    \gamma e\Bigg]\Phia(\Rg,0) \nonumber \\
  &\quad -\delta_{|j|1}\frac{\Rg}{2}e\ddp{\Phia}{R}(\Rg,0).
\end{align}
Here the guiding radius $\Rg(J_\phi)$ is defined by
$\Rg^2\Omega(\Rg)=J_\phi$, the eccentricity by $e(J_R,J_\phi) =
\sqrt{2J_R/(\kappa\Rg^2)}$, and $\gamma =
2\Omega(\Rg)/\kappa(\Rg)$.\footnote{In the potential from
  \Eq{eq:phi0}, $\kappa=\sqrt{2}\Omega$, so $\gamma=\surd2$.} In the
epicyclic approximation, the radial and azimuthal frequencies of an
orbit of actions $(J_R,J_\phi)$ are $\Omega_R=\kappa(\Rg)$, and
$\Omega_\phi=\Omega(\Rg) + [\ldd{\kappa(\Rg)}{J_\phi}]J_R$.

At a resonance, the orbital frequencies $\Omega_R$ and $\Omega_\phi$
satisfy
\begin{equation}
  l\Omega_R+m(\Omega_\phi-\Omb)=0.
\label{eq:resonance}
\end{equation}
The three main resonances are the corotation resonance ($l=0$), and
the outer ($l=1$) and inner ($l=-1$) Lindblad resonances. To capture
the behaviour of the slow and fast varying motions near the
resonances, one makes a canonical transformation of coordinates
defined by the following time-dependent generating function of type 2
\citep[e.g.,][]{Weinberg1994}
\begin{equation}\label{eq:S}
  S (\theta_R , \theta_\phi, \Js ,
  \Jf,t)=\paresq{l\theta_R+m\pare{\theta_\phi-\Omb t}}\Js+\theta_R\Jf.
\end{equation}
The new angles and actions $(\thf,\ths,\Jf,\Js)$ are then related to the
old ones by
\begin{equation}\label{eq:can}
\begin{aligned}
  \ths & =l\theta_R+m\pare{\theta_\phi-\Omb t},\quad & J_\phi & =m \Js, \\
  \thf &= \theta_R, \quad & J_R &= l\Js+\Jf.
\end{aligned}
\end{equation}

By taking the time-derivative of $\ths$ in the unperturbed system, and
by the definition of the resonance in Eq.~\ref{eq:resonance}, one
finds that the evolution of the new ``slow'' angle $\ths$ is indeed
slow near a resonance.  Given that along nearly circular orbits
$\theta_\phi\simeq\phi$, from \Eq{eq:can} we can understand why $\ths$
represents the azimuth of the apocentres and
pericentres\footnote{According to the convention of \cite{Dehnen1999},
  in this work the angle $\theta_R=0$ at the pericentre, and
  $\theta_R=\upi$ at the apocentre.} of the orbit in the frame of
reference that corotates with the bar: at $\theta_R=0$
($\theta_R=-\upi$), we are at the pericentre (apocentre) of the orbit
and $\ths$ ($\ths+\upi$) is $m$ times the star's azimuth
$\theta_\phi-\Omb t$ in the frame of reference corotating with the
orbit.

In the new canonical coordinates of \Eq{eq:can}, the motion in the
perturbed system is described by the following Hamiltonian (also
called the Jacobi integral)
\begin{equation}
  H=H_0+\Rep\parec{\sum_{j=-1}^1c_{jm}\eexp^{\img\paresq{(j-l)\thf+\ths}}}-m\Omb\Js,
\end{equation}
where $H_0$ is the Hamiltonian of the unperturbed axisymmetric system
and the $c_{jm}$ coefficients are the Fourier coefficients from
\Eq{eq:cR}, expressed as functions of the actions $(\Jf,\Js)$, thanks
to the canonical transformation from \Eq{eq:can}. Since $\thf$ evolves
much faster than $\ths$, we average
$H$ along $\thf$ \citep[the averaging principle,
  e.g.,][]{Arnold,Weinberg1994,BT2008}, to obtain
\begin{equation}
  \oH=H_0(\Jf,\Js)-m\Omb\Js+\Rep\parec{c_{lm}(\Jf,\Js)\eexp^{\img \ths}}.
\end{equation}
Since $\dot{\Jf}=-\lddp{\oH}{\thf}=0$, $\Jf$ is an integral of motion,
and for each $\Jf$, the motion of every orbit can be described purely
in the $(\ths,\Js)$ plane.

For each value of $\Jf$, let us then define $\Jsres$ as the value of $\Js$ where
\begin{equation}
  \Oms(\Jf,\Jsres)=0,
\end{equation}
where $\Oms \equiv l\Omega_R+m(\Omega_\phi-\Omb)$. While we expand
$H_0-m\Omb\Js$ in a Taylor series of $\Js$ around $\Jsres$ up to the
second order, we estimate $c_{lm}$ at $\Jsres$. Dropping the constant
terms, we obtain the approximate Hamiltonian near the resonances
\citep{Chirikov1979,Kaasalainen1994},
\begin{equation}\label{eq:Hpend}
  \widehat{H}=\frac{1}{2}G\pare{\Js-\Jsres}^2-F\cos\pare{\ths+g},
\end{equation}
where
\begin{equation}
  F \equiv -|c_{lm}\pare{\Jf,\Jsres}|, \quad G \equiv
  \ddp{\Oms}{\Js}\pare{\Jf,\Jsres},
\end{equation}
and $g\equiv\arg\pare{c_{lm}(\Jf,\Jsres)}$. \Eq{eq:Hpend} is the
Hamiltonian of a pendulum, and the equations of motion are
\begin{equation}\label{eq:eqmot}
\begin{aligned}
  \dot{\ths} &= G\pare{\Js-\Jsres}, \\
  \dot{\Js}  &= -F\sin\pare{\ths+g}. 
\end{aligned}
\end{equation}
Combining them, we obtain the equation for the $\ths$ acceleration,
namely,
\begin{equation}\label{eq:acc2}
  \ddot{\ths} = -\omega_0^2\sin\pare{\ths+g},
\end{equation}
where $\omega_0^2 \equiv FG$ (notice that in galaxies both $F$ and $G$
are negative).

The energy of the pendulum from \Eq{eq:acc2} is
\begin{equation}\label{eq:Ep}
  \Ep=\frac{\dot{\ths}^2}{2}+\Vp(\ths),
\end{equation}
where
\begin{equation}\label{eq:Vp}
  \Vp(\ths)=-\omega_0^2\cos(\ths+g).
\end{equation}
We can define the dimensionless quantity related to the energy
\begin{equation}
  k=\sqrt{\frac{1}{2}\pare{1+\frac{\Ep}{\omega_0^2}}}.
\end{equation}
For $k<1$, the orbit is trapped and librates around $\ths=-g$. In this
regime, the solution of \Eq{eq:eqmot} is \citep[e.g.,][]{Lawden1989}
\begin{equation}\label{eq:ths_lib}
  \ths+g=2\arcsin\pare{k~\sn\pare{\omega_0 t + C, k^2}},
\end{equation}
\begin{equation}\label{eq:Js_lib}
  \Js-\Jsres = \Ja~\cn\pare{\omega_0 t + C, k^2},
\end{equation}
where $\Ja \equiv 2\sqrt{F/G}k$, and $C$ the phase of the orbit. The
frequency of the oscillations of the librating pendulum is
\begin{equation}
  \omega = \omega_0 \frac{\upi}{2\mathrm{K}(k^2)}.
\end{equation}
The Jacobi functions $\sn$, $\cn$, and $\mathrm{K}$ are defined in
\App{app:Jacobi}.  Up to the second order, the expansion of
\Eqs{eq:ths_lib}{eq:Js_lib} in $k \ll 1$ leads to a solution
equivalent to that of an harmonic oscillator
\begin{equation}\label{eq:harmonicths}
  \ths+g \approx 2k~\sin\pare{\omega_0 t + C},
\end{equation}
\begin{equation}\label{eq:harmonicJs}
  \Js-\Jsres \approx \Ja~\cos\pare{\omega_0 t + C},
\end{equation}
with frequency $\omega \approx \omega_0$.

In the  circulating case, $k>1$, the solution of \Eq{eq:eqmot} for $\Js$
is
\begin{equation}\label{eq:Js_circ}
  \Js-\Jsres = \Ja~\dn\pare{\omega_0 k t + C, 1/k^2},
\end{equation}
where the Jacobi $\dn$ function is also defined in
\App{app:Jacobi}. While in this case $\ths$ is a monotonic function of
time, $\Js$ is an oscillating function of time around
$\langle\Js\rangle$. For $k\gg1$
\begin{equation}
  \Js \approx \Jsres + \Ja,
\end{equation}
which means that the angular momentum ($J_\phi=m\Js$) is conserved
very far from the resonance, i.e. we recover the axisymmetric case.

\begin{figure*}
  \centering 
  \includegraphics[width=0.9\columnwidth]{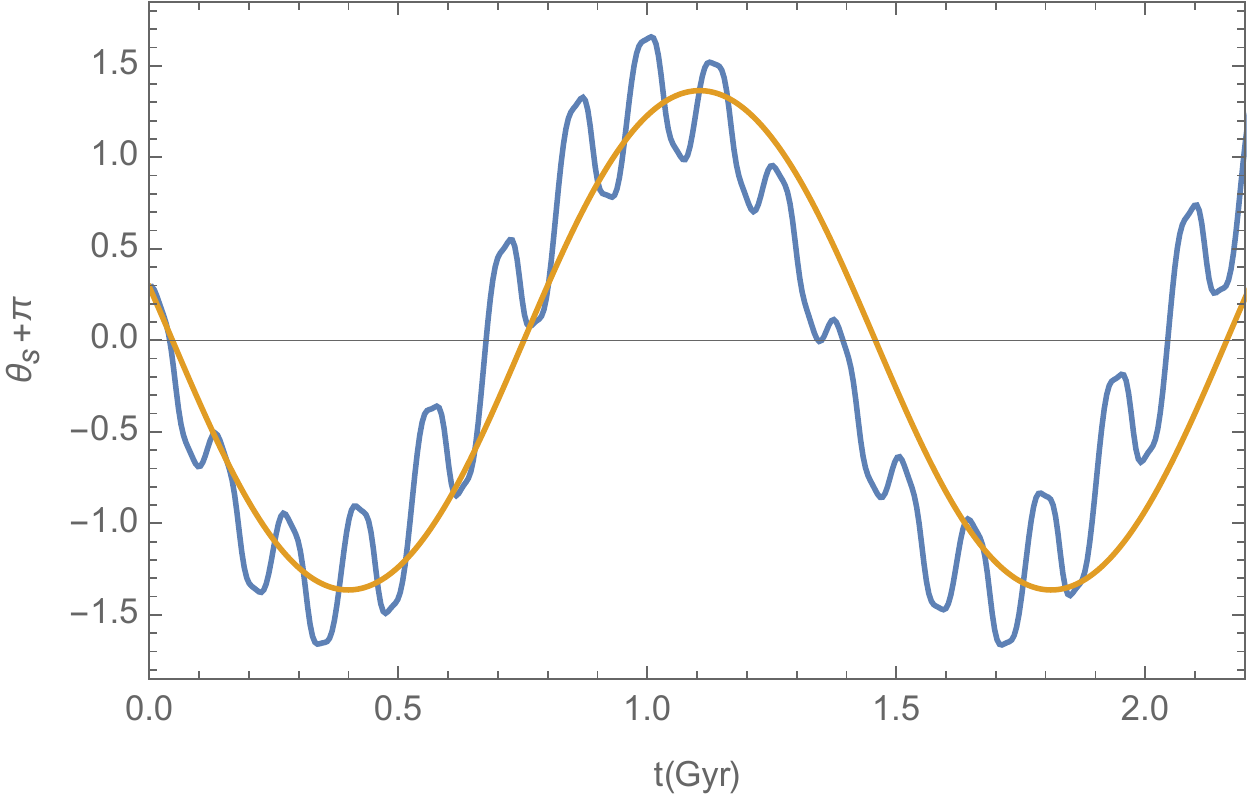}
  \includegraphics[width=0.9\columnwidth]{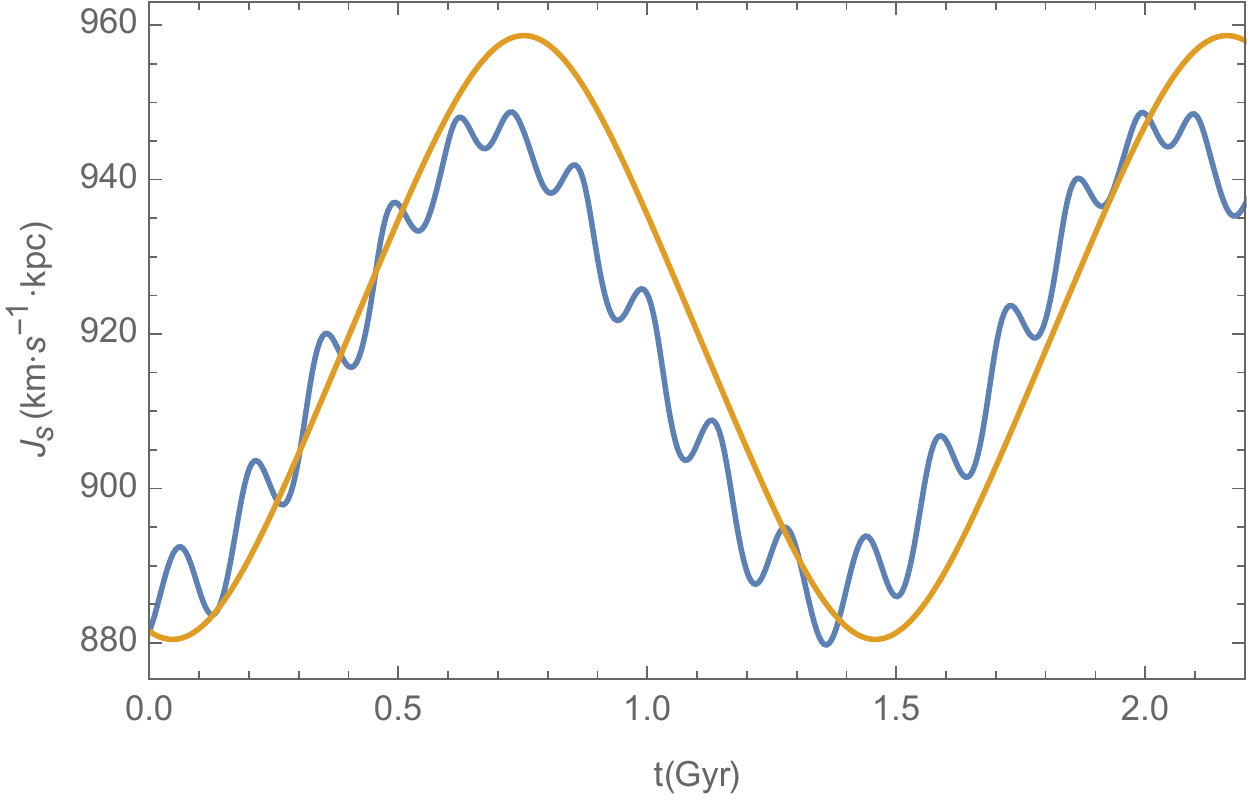}
  \caption{The slow variables $\ths+\upi$ (left panel) and $\Js$
    (right panel) for the librating orbit of \Fig{fig:trap_orb}. The
    blue lines correspond to the numerically integrated orbit, the
    orange lines to the pendulum approximation.}
\label{fig:thsJs}
\end{figure*}

As an example, in \Fig{fig:thsJs}, we follow the evolution of
$\ths+\upi$ (since $g=\upi$ in the case of the bar) and $\Js$ with
$(l,m)=(1,2)$ for the trapped orbit of \Fig{fig:trap_orb}.
The blue lines in these plots correspond to the orbit integrated
numerically. We see that the motion in $\ths$ and $\Js$ is a
composition of high frequency, low amplitude oscillations (that are
ignored, when invoking the averaging principle), and slow frequency
high amplitude oscillations. The pendulum approximation (orange
lines), provides a description of the latter.\footnote{A more accurate
  description of the orbit can be obtained by performing a limited
  development at higher order \citep{Binney2016}.}

\section{Actions and angles for the pendulum}

The action and the angle associated with the pendulum are
\citep[e.g.,][]{Brizard2013}, for the case $k<1$,
\begin{equation}
\begin{aligned}
  \Jp =&
  \frac{4}{\upi}\frac{\Ja}{k}\paresq{\mathrm{E}(k^2)-(1-k^2)\mathrm{K}(k^2)}, \\
  \thp =&\frac{\upi}{2\mathrm{K}(k^2)}\cn^{-1}\pare{\frac{\Js-\Jsres}{\Ja},k^2}.
\end{aligned}
\end{equation}
Using $\thp$ we can rewrite $\ths+g$ as
\begin{equation}
  \ths+g=2\arcsin\pare{k~\sn\pare{\frac{2}{\upi}\mathrm{K}(k^2)\thp,
      k^2}},
\end{equation}
and 
\begin{equation}\label{eq:Jscos}
  \Js-\Jsres=\Ja~\cn\pare{\frac{2}{\upi}\mathrm{K}(k^2)\thp,
      k^2}.
\end{equation}
For the case $k>1$, one has 
\begin{equation}
\begin{aligned}
  \Jp  =&
  \frac{2}{\upi}\Ja\mathrm{E}(1/k^2),\\
  \thp =&\frac{\upi}{\mathrm{K}(1/k^2)}\dn^{-1}\pare{\frac{\Js-\Jsres}{\Ja},1/k^2}.
\end{aligned}
\end{equation}
We can rewrite 
\begin{equation}\label{eq:JsJa}
  \Js-\Jsres=\Ja~\dn\pare{\frac{\mathrm{K}(1/k^2)}{\upi}\thp, 1/k^2}.
\end{equation}

\begin{figure}
  \centering 
  \includegraphics[width=0.95\columnwidth]{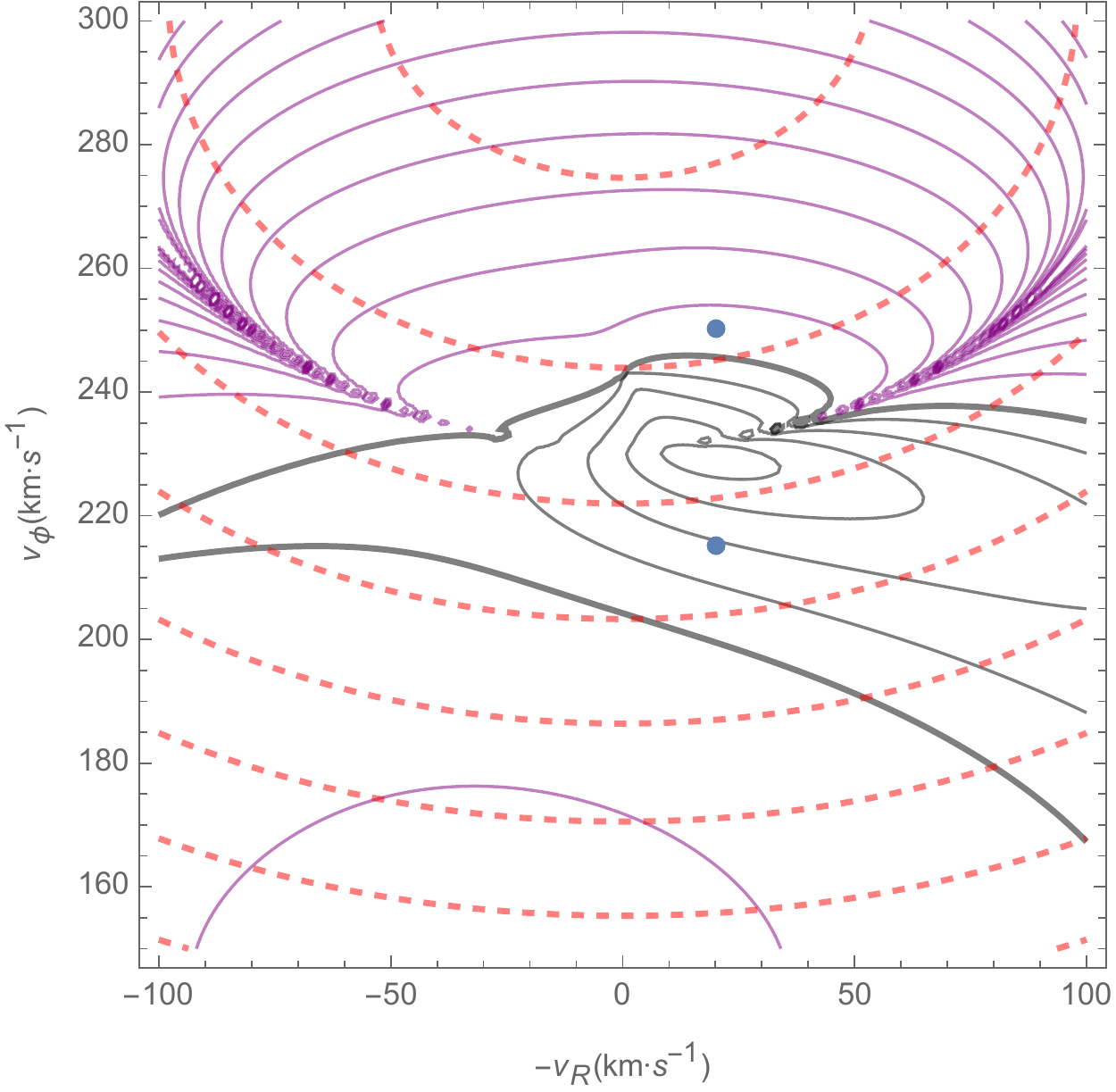}
  \caption{Contours of conserved quantities associated with the pendulum
    treatment at the outer Lindblad resonance in the $(-v_R,v_\phi)$
    plane at $(R,\phi)=(8\Kpc,-25\degr)$ for the bar model presented
    in this work with $\Omb=1.8\Omega_0$. The black contours
    correspond to constant values of $k$ for $k\leq 1$. The thick
    black contour corresponds to $k=1$. The purple contours correspond
    to constant values of $k$ for $k>1$. The red dashed contours
    correspond to constant values of $\Jf$. The blue dots correspond
    to the initial conditions of the orbits shown in
    \Fig{fig:trap_orb}.}
  \label{fig:k}
\end{figure}
In \Fig{fig:k}, we show $k$ and $\Jf$ contours in the velocity
space\footnote{The minus in front of $v_R$ is chosen to allow a better
  comparison with the data of kinematics of stars in the Galaxy,
  usually plotted in the $(U,V)$ space, with $U$ positive towards the
  Galactic centre.}  $(-v_R,v_\phi)$, for $(R,\phi)=(8\Kpc,-25\degr)$
with the same potential used to integrate the orbits in
\Fig{fig:trap_orb}. The black contours are for $k<1$ (trapped orbits),
the green contours for $k>1$ (circulating orbits). The red dashed
contours correspond to contours of constant $\Jf$. The quantities $k$
(or $\Jp$) and $\Jf$ characterize an orbit.\footnote{The quantity
  $\Jsres$ is fixed by $\Jf$, from the condition
  $\Oms(\Jf,\Jsres)=0$.} The blue points correspond to the initial
conditions of the two orbits in \Fig{fig:trap_orb}, which both start
from $(R,\phi)=(R_0,-25\degr)$. The minimum $k=0$ is found at
$(-v_R,v_\phi)\approx(25,230)\kmsec$ and corresponds to the most
trapped orbit at the outer Lindblad resonance for such $(R,\phi)$.

\section{Averaging distribution functions over pendulum angles}

Following the prescription of \cite{Binney2016}, we assume that the
distribution function for orbits trapped by the resonances ($k<1$) at
a certain point $(\Jf,\Js,\ths)$ is given by the average of the
unperturbed DF $f_0$ along $\thp$, i.e.,
\begin{equation}\label{eq:int}
  \ftr(\Jf,\Js,\ths)=\frac{1}{2\upi}\int_0^{2\upi}
    f_0(\Jf,\Jsres+\Delta\Js(\thp)) \de\thp,
\end{equation}
where from \Eq{eq:JsJa}
\begin{equation}\label{eq:DJs}
  \Delta\Js \equiv \Ja~\cn\pare{\frac{2}{\upi}\mathrm{K}(k^2)\thp,
    k^2},
\end{equation}
and $k$ is a function of $(\Jf,\Js)$, and $\ths$, derived from $R$,
$\phi$, $v_R$, $v_\phi$, and the potential. The physical meaning of
\Eq{eq:int} is that $\ftr$ corresponds to the unperturbed distribution
function $f_0$ phase-mixed along the pendulum angle, assuming that
enough time elapsed since the growth of the perturbation.

The value of the integral in \Eq{eq:int} depends on the particular
form of $f_0$. Therefore, in general its solution can be computed
numerically as
\begin{equation}
  \ftr(\Jf,\Js,\ths)=\frac{1}{N}\sum_i
    f_0(\Jf,\Jsres+\Delta\Js(\thp^i)),
\end{equation}
where $\thp^i$ sample the orbit between $0$ and $2\upi$, and $N$ is
the number of sampling points.

For $k \ll 1$, we can even give an analytic form for the distribution
function. To solve this integral, we expand $\ln(f_0)$ around $\Jsres$
(in typical galaxies $f_0$ is almost exponential in $\Js$) as
\begin{equation}
  \ln\pare{f_0(\Jf,\Jsres+\Delta\Js)} \approx
  \ln\pare{f_0(\Jf,\Jsres)}-\frac{\Delta\Js}{\Jh},
\end{equation}
where
\begin{equation}
  \Jh \equiv -1\bigg/\pare{\frac{1}{f_0}\ddp{f_0}{\Js}}_{\Jsres}.
\end{equation}
Then
\begin{equation}\label{eq:f0app}
  f_0(\Jf,\Jsres+\Delta\Js) \approx f_0(\Jf,\Jsres)\eexp^{-\Delta\Js/\Jh}.
\end{equation}
This approximation is excellent to express $f_0$, but unfortunately,
it is not enough to solve the integral from \Eq{eq:int}. To do that
one more approximation is necessary. Expanding \Eq{eq:DJs} to first
order in $k$, $\Delta\Js$ can be expressed as
\begin{equation}\label{eq:Jsapp}
  \Delta\Js\approx \Ja \cos{\thp},
\end{equation}
In this way, {\it de facto}, we fall back on the harmonic oscillator solution
\Eqs{eq:harmonicths}{eq:harmonicJs}. With these approximations, the solution
of the integral from \Eq{eq:int} is
\begin{equation}\label{eq:ftr}
  \ftr(\Jf,\Js,\ths)=f_0(\Jf,\Jsres)\mathrm{I}_0(\Ja/\Jh),
\end{equation}
where $\mathrm{I}_0(x)$ is the incomplete Bessel function of the first
kind.

For the zone of circulation ($k>1$) we instead use for the
distribution function the prescription
\begin{equation}
  \fcirc(\Jf,\Js,\ths)=f_0(\Jf,\overline{\Js}),
\end{equation}
where  
\begin{equation}
  \overline{\Js} = \frac{1}{2\upi}\int_0^{2\upi}\Js(\thp) \de\thp =
  \Jsres + \frac{\upi}{2}\frac{\Ja}{K(1/k^2)}.
\end{equation}
 This prescription is motivated by the fact that, outside
of the trapping region, the perturbing potential simply deforms the orbital
tori of the underlying axisymmetric system rather than abruptly building
completely new tori as it does within the trapping region. Consequently, if
the perturbation emerges slowly enough, the phase-space density will be
adiabatically constant on each torus as it is deformed at its original value,
$f_0(\bJ)$, where $\bJ=(\Jf,\overline{\Js})$ is to be understood to be the invariant actions of
the perturbed torus rather than momenta of the original system of
angle-action variables.
Notice that, for large $k$ (far from the
resonance), $\overline{\Js} \rightarrow \Js$ rather fast, and we are
back to the axisymmetric case (conservation of the angular momentum
$J_\phi=m\Js$).

\section{Results}

We now present a few results of astrophysical interest for the
response at the resonances of an unperturbed distribution function
$f_0$ to the bar perturbation presented in the previous sections.

As an unperturbed distribution function $f_0$ we choose
\citep{Binney2011}
\begin{equation}
 f_0(J_R,J_\phi)=\frac{\gamma(\Rg)\Sigma_0\eexp^{-\Rg/h_R}}{2\upi\sigma_R^2(\Rg)}
 \eexp^{-\frac{J_R\kappa(\Rg)}{\sigma_R^2(\Rg)}},
\end{equation}
where
\begin{equation}
 \sigma_R(R)=\sigma_R(R_0)\eexp^{-\frac{R-R_0}{5 h_R}},
\end{equation}
with $\gamma$ defined as in \Eq{eq:cR}, $\sigma_R(R_0)=30\kmsec$, and
$h_R=2\Kpc$, a reasonable description of the kinematics of the Solar
neighbourhood.

\begin{figure*}
 \centering
 \includegraphics[width=0.95\textwidth]{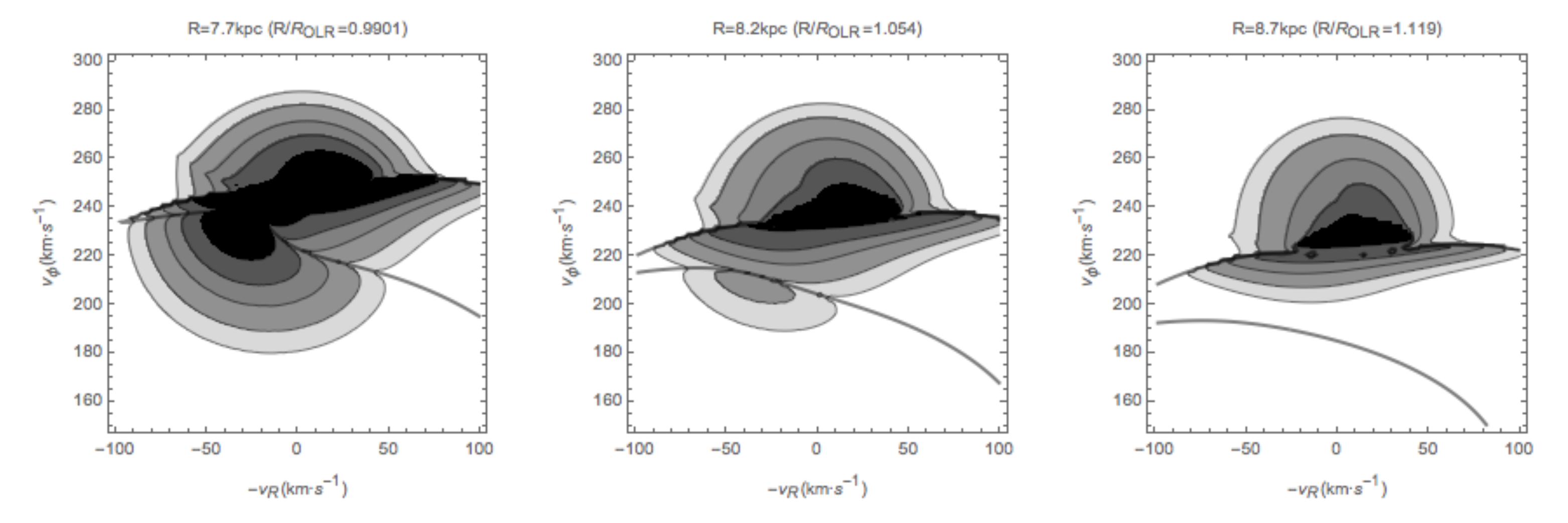}
 \caption{Distribution function in the $(-v_R,v_\phi)$ plane at
   different Galactocentric radii $R$ and $\phi=-25\degr$, obtained
   using the treatment at the resonances and the bar model presented
   in this work. The bar's pattern speed is $\Omb=1.8.\Omega_0$. The
   thick contours enclose orbits trapped at the outer Lindblad
   resonance. The distribution function contours enclose (from the
   outer to the inner) 95, 90, 80, 68, and $50\%$ of the stars.}
 \label{fig:DFOLR}
\end{figure*}
We first consider, in \Fig{fig:DFOLR}, the density of stars in local
velocity space obtained from a model with a ``fast'' rotating bar, as
in the classical picture
\citep{Dehnen2000,Antoja2014,Monari2017b}. For such a bar, the Solar
neighbourhood is in the vicinity of the outer Lindblad resonance of
the bar. The angle of the bar with respect to the Solar position is
taken to be $\phi=-25\degr$.  The bar that we present here has
$\Omb=1.8\Omega_0$. We see that, in line with previous studies, the
analysis in this work also predicts the formation of a low-velocity
overdensity similar to the Hercules moving group
\citep[e.g.,][]{Dehnen1998,Famaey2005} at positive $v_R$, whose
velocity position and relative amplitude varies as a function of
radius. The group is not formed by orbits trapped by the outer
Lindblad resonance, but from circulating orbits with guiding radii
inside the $\ROLR$. The orbits trapped to a resonance rather seem to
be associated with the feature of local velocity space sometimes
called the ``horn'' \citep[e.g., ][]{MonariPhD}. This is also in line
with previous studies, but never before had the distribution function
in the trapped region been quantified for a fully phase-mixed
population. Interestingly, we clearly see that in this case the
Hercules moving group shifts both in $v_\phi$ (lower $v_\phi$ at
larger radii) and in $-v_R$ (larger $-v_R$ at larger radii).

\begin{figure*}
 \centering
 \includegraphics[width=0.95\textwidth]{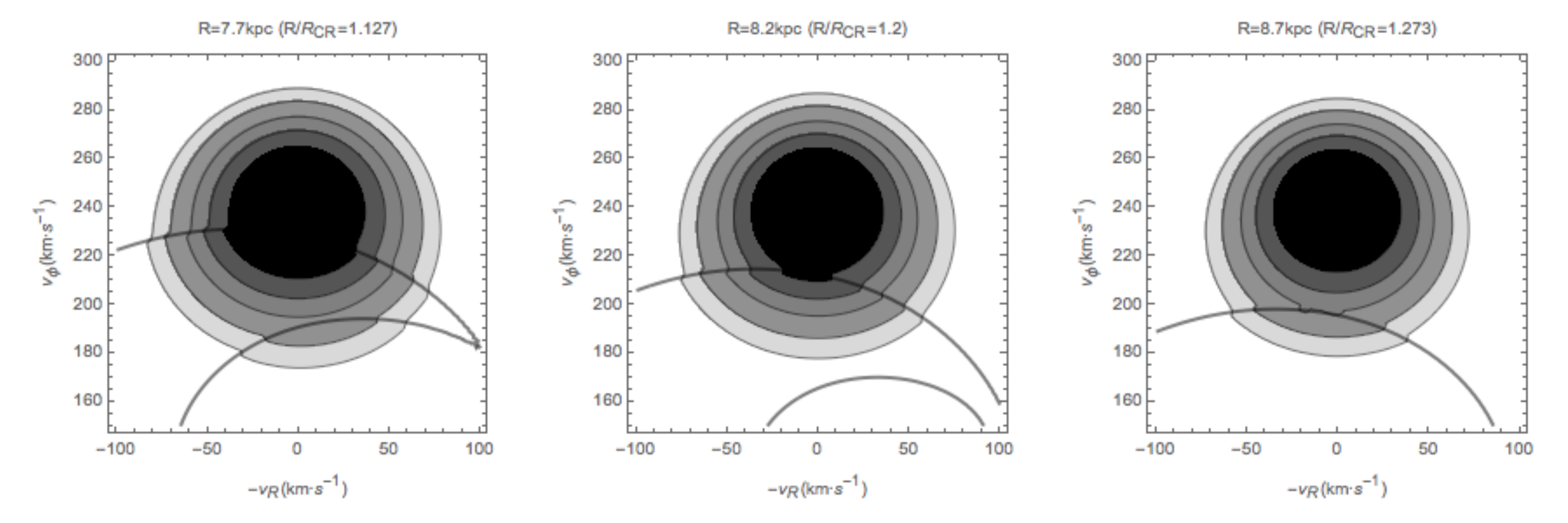}
 \caption{As in \Fig{fig:DFOLR}, but for a bar's pattern speed
   $\Omb=1.2\Omega_0.$ The thick contours enclose orbits trapped to the
   corotation.}
 \label{fig:DFCR}
\end{figure*}
\begin{figure}
 \centering
 \includegraphics[width=0.95\columnwidth]{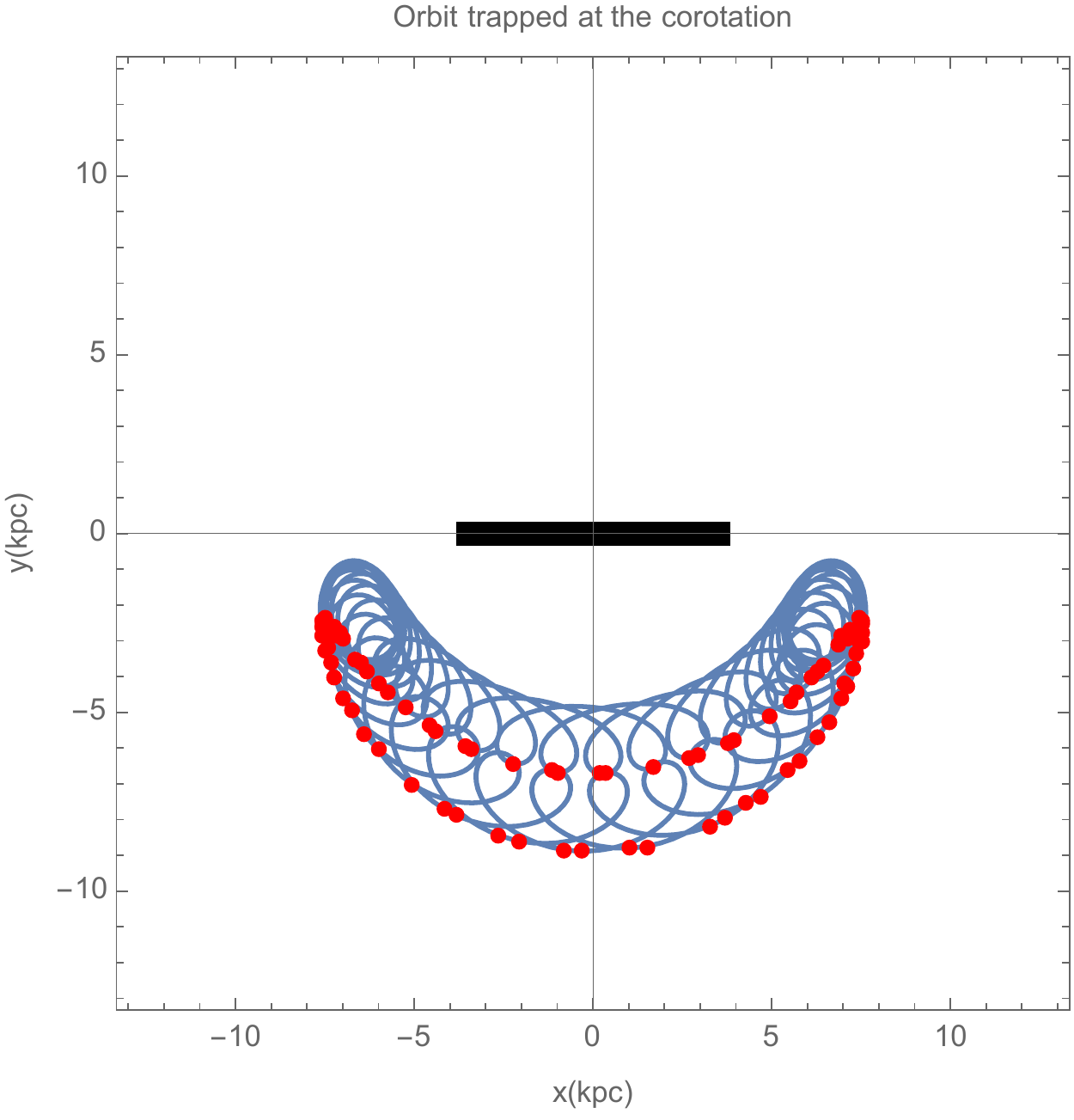}
 \caption{An orbit trapped to the corotation, integrated in the
   potential presented in this work with a bar's pattern speed
   $\Omb=1.2\Omega_0$. The orbit is shown in the reference frame
   corotating with the bar. The initial conditions of this orbit are
   $(R,\phi)=(8.2\Kpc,-25\degr)$, and
   $(-v_R,v_\phi)=(-10,210)\kmsec$. The red dots correspond to the
   position of the apocentres, and the black thick line to the long
   axis of the bar. The bar rotates counterclockwise.}
 \label{fig:Lag}
\end{figure}
\begin{figure*}
 \centering
 \includegraphics[width=0.95\textwidth]{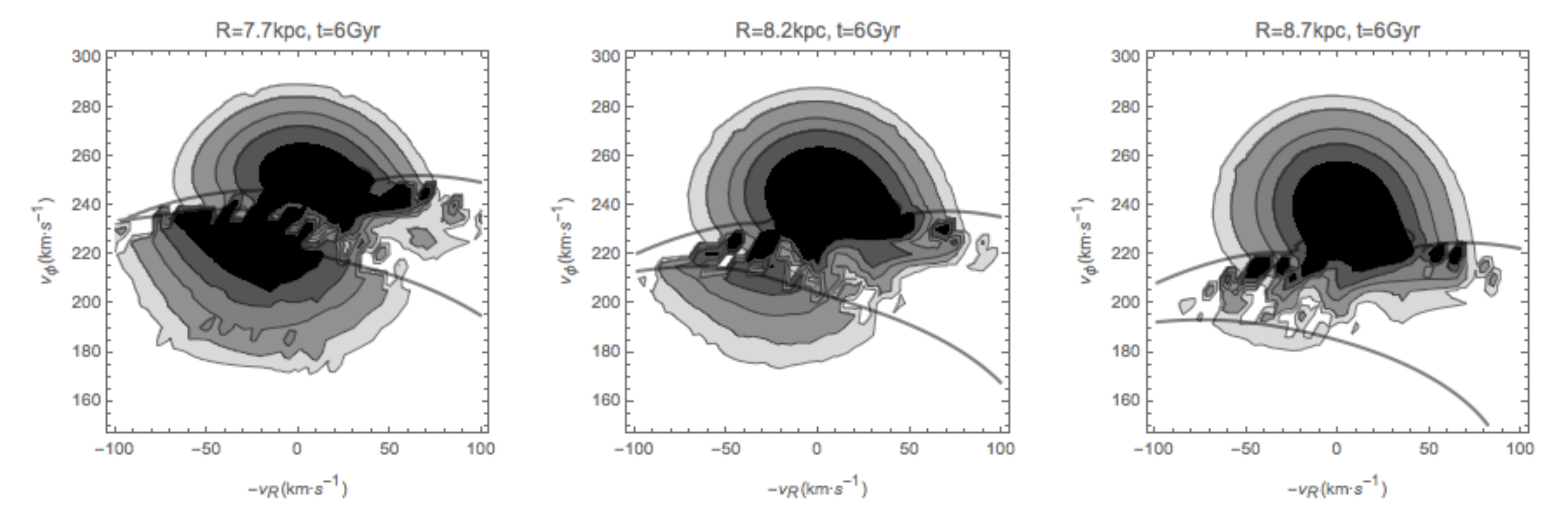}
 \includegraphics[width=0.95\textwidth]{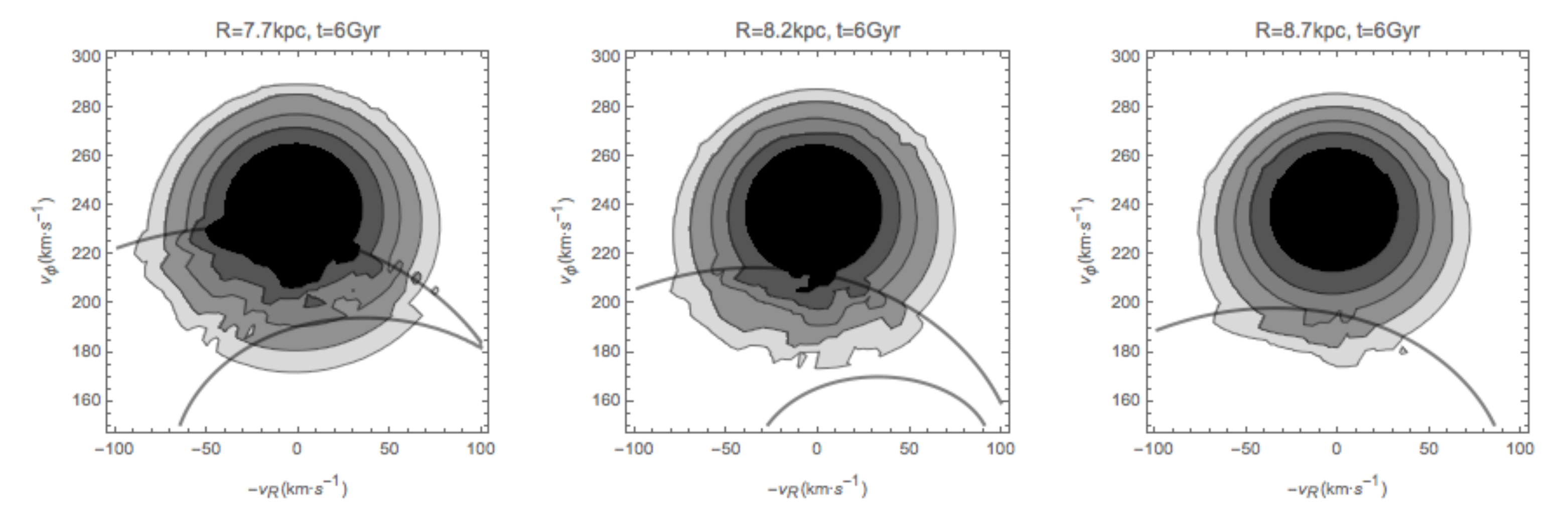}
 \caption{Local velocity distribution obtained from numerically
   integrating orbits for 6 Gyr after an adiabatic growth of the bar,
   starting from the same initial $f_0$ distribution function as for
   \Fig{fig:DFOLR} and \Fig{fig:DFCR}. The resolution of the grid used
   to obtan this figure is $5\kmsec$. Compare the top-row ($\Omb = 1.8
   \Omega_0$) with \Fig{fig:DFOLR}, and the bottom row ($\Omb = 1.2
   \Omega_0$) with \Fig{fig:DFCR}.}
 \label{fig:DFint}
\end{figure*}

Recently, \cite{Sormani2015} and \cite{Li2016} have argued that the
pattern speed of the bar is $\Omb=1.2\Omega_0$, significantly lower
than in the classical picture. In this case the solar neighbourhood
would lie just outside corotation, so in \Fig{fig:DFCR} we also plot
the velocity distribution at three such locations. In the central and
right panels of \Fig{fig:DFCR}, when the zone of trapping is at low
azimuthal velocities, a deformation in the velocity distribution at
negative $-v_R$ that could resemble Hercules forms {\it within} the
trapping zone rather than outside it. When there is a region of
enhanced density below the trapping zone, (left panel of
\Fig{fig:DFCR}) it occurs at $-v_R>0$, as predicted by the {\it
  Eulerian} linear theory, and as such conflicts with the
observations. Hence, the pendulum formalism is mandatory if one seeks
to explain the Hercules group as a consequence of the corotation
resonance, as \cite{Villegas2017} did with made-to-measure $N$-body
models. Notice that the orbits associated with the overdensity at
$(-v_R,v_\phi) \approx (-10,210) \kmsec$ in the central panel of
\Fig{fig:DFCR} are of the same kind of those described by
\cite{Villegas2017}, i.e., trapped around the bar's Lagrangian points
(see \Fig{fig:Lag}).
Finally we provide the reader with a comparison plot in
\Fig{fig:DFint}, in which we display the local velocity distribution
obtained from orbits integrated for 6 Gyr in the same potential, after
the bar is slowly grown for 3 Gyr with the growth law from
\cite{Dehnen2000}, and for the same initial $f_0$ as in our analytic
model \citep[with the backwards integration method
  of][]{VauterinDejonghe1997,Dehnen2000}.

\section{Conclusion}

In this paper, we presented for the first time a way to treat an
action-based distribution function in the region of action space where
orbits are resonantly trapped by a bar, where the {\it Eulerian}
treatment of \citet{Monari2016} diverges. The idea is rather to follow
the deformation of the tori outside the trapping region, while
averaging the distribution function over the relevant angles in the
trapping region. We showed that in the trapping region the relevant
action-angle variables are those of a pendulum, and averaging over
those angles allows for a smooth connection with the deformed tori in
the circulation zone. With such distribution functions, we can
reproduce an overdensity in velocity space resembling the Hercules
moving group both outside the outer Lindblad resonance and outside
corotation of a bar.  The linearized Eulerian treatment of
\cite{Monari2017a} is unable to handle the latter possibility. The
disturbances in velocity space that are caused by the inner Lindblad
and corotation resonances move in different ways through velocity
space as one changes location within the disc (Figs~\ref{fig:DFOLR}
and \ref{fig:DFCR}). Consequently, it will be straightforward to
determine which resonance is responsible for the Hercules group, once
we can reliably map velocity space in many locations.

This formalism opens the way to fitting quantitatively the effects of the bar
in an action-based modelling of the Milky Way. Nevertheless, there remain
multiple tests to be done, which are beyond the scope of the present
contribution presenting the relevant formalism. In particular, the prediction
of a fully phase-mixed distribution function in the trapping region should be
compared to the outcome of various simulations, to check over which
time-scales phase-mixing is efficient enough to reproduce our results.
Moreover, the process of trapping and the associated filling of the region of
resonant trapping in action space is not necessarily going to be based purely
on the phase-mixing of the original axisymmetric distribution function,
especially if the growth of the bar is rapid. But even in this case, the
advantage of the present paper has been to present the relevant pendulum
action variables on which to base a parametric distribution function to fit
both simulations and real data in the trapping zone.

As a matter of fact, only the upcoming Gaia data will allow us to
check whether our phase-mixing of the original distribution function
is actually a good representation of the Galactic disc at different
radii. If not, knowing that our distribution function must depend only
on the new set of action variables within the trapping region, we will
be able to leave its functional form free, and fit it to the
data. This approach is very fast and does not require to perform
numerous expensive simulations. As a consequence, one will be able to
explore very efficiently the parameter space of the perturbations.

Further improvements of the present formalism will need taking into
account the vertical $z$ direction, the time dependence in the
amplitude of perturbations, as well as collective effects
\citep[e.g.,][]{Weinberg1989, FouvryPichMagorChavani2015}. It will
also be mandatory to move away from the epicyclic approximation
\citep{torus,SandersBinney2015}. Once this will be done, in the
absence of strong resonance overlaps, a complete dynamical model of
the present-day Milky Way disc could then in principle finally be
built by applying, on top of the trapped distribution function near
the main resonances of each perturber, our previous {\it Eulerian}
treatment of perturbations~\citep{Monari2016} for the other
perturbers, even including vertical perturbations and ``bending''
modes of the disc \citep{Widrow2014,Xu2015,Laporte2016}.

\section*{Acknowledgements}
JBF acknowledges support from Program HST-HF2-51374 which was provided
by NASA through a grant from the Space Telescope Science Institute,
which is operated by the Association of Universities for Research in
Astronomy, Incorporated, under NASA contract NAS5-26555. This work was supported by
the European Research Council under the European Union's Seventh Framework
Programme (FP7/2007-2013)/ERC grant agreement no.~321067.

\bibliographystyle{mn2e} \bibliography{mn-jour,trappingbib}
\label{lastpage}

\appendix

\section{Jacobi special functions}\label{app:Jacobi}

The Jacobi $\sn$, $\cn$, and $\dn$ functions can be evaluated as the
sum of a power series, and are defined as
\begin{equation}
  \sn(u,m) \equiv \sin(\varphi), \quad
  \cn(u,m) \equiv \cos(\varphi),
\end{equation}
with
\begin{equation}
  u=\int_0^\varphi\frac{\de \theta}{\sqrt{1-m\sin^2\theta}}.
\end{equation}
The $\dn$ function is defined as
\begin{equation}
  \dn(u,m)^2 \equiv 1-m~\sn^2(u,m).
\end{equation}
The elliptic functions $\mathrm{E}$ and $\mathrm{K}$ are defined as
\begin{equation}
  \mathrm{E}(m)=\int_0^{\upi/2}\de \theta{\sqrt{1-m\sin^2\theta}},
\end{equation}
\begin{equation}
  \mathrm{K}(m)=\int_0^{\upi/2}\frac{\de \theta}{\sqrt{1-m\sin^2\theta}}.
\end{equation}

\end{document}